\documentstyle[preprint,eqsecnum,aps,floats]{revtex}
\input epsf
\begin{document}

\begin{titlepage}

\preprint{NSF-ITP-96-49, UCSBTH-95-28}

\title{Solutions of the Regge Equations on some Triangulations of $CP^2$} 
\tighten
\author{James B.~Hartle\thanks{Present address: Institute for Theoretical 
Physics, University of California, Santa Barbara, CA, 93108; e-mail 
address:
hartle@itp.ucsb.edu} and Zoltan Perj\'es\thanks{Permanent address: 
Central Research
Institute for Physics,
PO Box 49, H-1525 Budapest, HUNGARY;
e-mail address: perjes@rmk530.rmki.kfki.hu}} \vskip .26 in

\address{Department of Physics, University of California\\ Santa Barbara, 
CA 93106-9530}

\date{\today}

\maketitle

\begin{abstract}

Simplicial geometries are collections of simplices making up a manifold 
together with an assignment of lengths to the edges that define a metric 
on that manifold. The simplicial analogs of the Einstein equations are 
the Regge equations. Solutions to these equations define the 
semiclassical approximation to simplicial approximations to a 
sum-over-geometries in quantum gravity. In this paper, we consider 
solutions to the Regge equations with cosmological constant that give 
Euclidean metrics of high symmetry on a family of triangulations of 
$CP^2$ presented by Banchoff and K\"uhnel. This family is characterized 
by a parameter $p$. The number of vertices grows larger with increasing 
$p$. We exhibit a solution of the Regge equations for $p=2$ but find no 
solutions for $p=3$. This example shows that merely increasing the number 
of vertices does not ensure a steady approach to a continuum geometry in 
the Regge calculus. 

\end{abstract}

\end{titlepage}

\tighten

\section{Introduction}
\label{sec:I}

The sums over geometries that arise in quantum gravity may be 
approximated, and perhaps even defined, by the methods of the Regge 
Calculus.\footnote{The original paper is T.~Regge \cite{Reg61}. For a 
review and extensive list of references see Williams and Tuckey 
\cite{WT92}.} Smooth, four-dimensional, geometries are approximated by 
simplicial geometries built from a finite number of four-simplices joined 
together so as to give a triangulation of a manifold. Different manifolds 
are
represented by putting simplices together in different ways. Different 
metrics are obtained (in general) by different assignments of lengths to 
the edges of the simplices. A sum over topologies is approximated by a 
sum over different ways of assembling four-simplices. A sum over metrics 
on a given manifold is approximated by a multiple integral over the 
squared edge-lengths. In the Euclidean functional integral approach to 
quantum gravity typical integrals have the form
\begin{equation}
\int_C d\Sigma_1 A(s_i,M) {\rm exp}[-I(s_i,M)/\hbar]\ , \label{oneone}
\end{equation}
where $M$ is a closed, compact simplicial four-manifold, $I(s_i,M)$ is 
the Regge gravitational action, and $A(s_i,M)$ is some quantity of 
interest to be averaged in this way. Both $I$ and $A$ are functions of 
the squared edge-lengths $s_i, i=1,...,n_1$ which are integrated along a 
contour $C$ with an appropriate measure $d\Sigma_1$. Such simplicial 
approximations to sums over geometries have been discussed in several 
places \cite{Har85a,Hamsum} and applied to a number of problems in 
quantum cosmology \cite{Eli89,Mor94,Fur96}. 

In some limits, the integral (1.1) may be evaluated semiclassically using 
the method of steepest descents. The value of the integral is then 
dominated by the contribution near one or more of stationary points of 
the action (through which the steepest descents distortion of the contour 
passes). At these,
\begin{equation}
{{\partial I(s_j,M)}\over{\partial s_i}}=0, \quad\quad i=1,...,n_1\ . 
\label{onetwo}
\end{equation}
These algebraic equations are the simplicial analogs of the Einstein 
equations and are called the Regge equations. Solutions of the Regge 
equations on various triangulations of different manifolds are therefore 
of interest.

Large sets of algebraic equations like (1.2) are not always easy to 
solve, especially when the domain of the solution is constrained, as it 
is here because of the triangle inequalities and their higher dimensional 
analogs. However, when the triangulation has a symmetry, solutions 
consistent with that symmetry are more easily obtained. A number of 
solutions of this type on triangulations of the manifolds $S^4$ and 
$CP^2$ were exhibited in \cite{Har86a} (Paper II). Recently Banchoff and 
K\"uhnel \cite{BK92} have exhibited a family of symmetric triangulations 
of the manifold $CP^2$. With such families, it is possible to
observe the behavior of
the Regge approximation to a given spacetime as the number of vertices is 
increasing. In this paper we investigate the solution of the Regge 
equations (1.2) on some of these triangulations.

For completeness, some requisite
properties of $CP^2$ are summarized in Section II. The Banchoff-K\"uhnel 
triangulations are described briefly in Section III. They are 
characterized by an integer $p \ge 2$. With $n=p^2+p+1$, each 
triangulation
$CP^2_{n+3}$ has $p^2+p+4$ vertices, $3pn$ edges, $2(6p-5)n$ triangles, 
$15(p-1)n$ tetrahedra,
$6(p-1)n$ four-simplices, and a symmetry group of order $6n$.
Maximally symmetric solutions with edges related by this symmetry group 
are specified by $p$ independent edge-lengths. Thus, by varying $p$ one 
has an increasingly large family of highly symmetric triangulations of 
$CP^2$.

We use a computer program to find analytic expressions for the action 
$I(s_i,CP^2_{n+3})$ as a function
of the independent edges in a symmetric assignment of the edges. We 
exhibit a solution of the Regge equations (1.2) for $p=2$ and compare it 
with that of one other triangulation \cite{KL83} not in this family. For 
$p=3$
we find no solution. This shows that, even with triangulations of high 
symmetry, one cannot always expect to find the discrete analogs of the 
continuum solutions of Einstein's equations. We discuss the reason this 
family of triangulations exhibits these phenomena.

\section{The Fubini-Study Metric of $CP^2$} \label{sec:II}

The complex projective plane $CP^2$ is defined by equivalence classes of 
points $(Z^1,Z^2,Z^3)$
of the complex Euclidean 3-space $C^3$ with the origin excluded. The 
equivalence relation is $(Z^1,Z^2,Z^3) = (\lambda Z^1,\lambda Z^2,\lambda 
Z^3)$ for any non-zero complex $\lambda$. $CP^2$ possesses a continuum 
metric of high symmetry called the Fubini-Study metric. We follow Gibbons 
and Pope \cite{GP78} in a brief review of its properties. 
Points in
$CP^2$ may be labeled by co\"ordinates
\begin{equation}
\zeta^i=Z^i/Z^3,\qquad i=1,2
\label{twoone}
\end{equation}
for $Z^3\not= 0$.

As described in \cite{GP78}, the Fubini-Study metric on $CP^2$ is 
constructed from the
Euclidean metric on the round 5-sphere in ${\bf C}^3$, \begin{equation}
|Z^1|^2+|Z^2|^2+|Z^3|^2={6\over\Lambda}\ , \label{twotwo}
\end{equation}
where $\Lambda $ is a constant.
Defining Euler angles $(\psi,\theta,\phi)$ and a radial coordinate $r$ by 
\begin{equation}
\zeta^1=r\cos(\theta/2)\, e^{i(\psi+\phi)/2}\ , \label{twothree}
\end{equation}
\begin{equation}
\zeta^2=r\sin(\theta/2)\, e^{i(\psi-\phi)/2}\ , \label{twofour}
\end{equation}
the Fubini-Study metric is
\begin{equation}
ds^2={dr^2\over(1+\lambda r^2)^2}+{r^2\over4(1+\lambda 
r^2)^2}(d\psi+\cos\theta d\phi)^2 +{r^2\over4(1+\lambda 
r^2)}(d\theta^2+\sin^2\theta d\phi^2) \label{twofive}
\end{equation}
where $\lambda=\Lambda/6$.
This coordinate system for
$0\leq\theta\leq\pi,0\leq\phi\leq2\pi,0\leq\psi\leq4\pi$ and $0\leq 
r\leq\infty$ covers ${\bf R}^4$ except at $r=0,\theta=0$ and 
$\theta=\pi$. 

The Euclidean metric (\ref{twofive}) satisfies the Einstein equation 
\begin{equation}
R_{\alpha\beta} = 4\Lambda g_{\alpha\beta} \label{twosix}
\end{equation}
so that $\Lambda$ is the cosmological constant. It is to this solution 
of the continuum Einstein equations that we shall compare our solutions 
of the Regge equations on triangulations of $CP^2$. 

\section{The Triangulation}
\label{sec:III}

A remarkable sequence of triangulations of the complex projective plane 
$CP^2$ has been presented by Banchoff and K\"uhnel \cite{BK92}. It is 
based on Coxeter's regular map $\{3,6\}_{1,p}$ of the flat torus $T^2$. 
The latter is defined by an array of \begin{equation}
n=p^2+p+1
\label{threeone}
\end{equation}
vertices (where $p=2,3,...$) of a
tessellation of the two-torus $T^2$ by equilateral triangles. Numbering 
these vertices by the non-negative integers $k\ (mod\ n)$, they may be 
located on a two dimensional ``plane'' in $CP^2$ at values of the 
co\"ordinates (\ref{twoone}) \begin{equation}
\zeta^1=e^{2\pi ik/n},\qquad\zeta^2=e^{-2\pi ipk/n}\ . \label{threetwo}
\end{equation}
The triangulation of $CP^2_{n+3}$ is constructed as the union of three 
solid 4-balls $B_1,B_2,B_3$. The flat torus $T^2$ is the common 
intersection of the three balls. The triangulation of each of these balls 
consists of 4-simplices, four
vertices of which lie on the boundary torus $T^2$ and one vertex at the 
point of $CP^2$ defined by the following points in $C^3$: 
\begin{eqnarray}
X=(1,0,0)\qquad	& {\rm for} \ B_1\ ,\nonumber\\
Y=(0,1,0)\qquad	& {\rm for} \ B_2\ ,\nonumber\\
Z=(0,0,1)\qquad	& {\rm for} \ B_3\ .
\label{threethree}
\end{eqnarray}
Three kinds of chains of 4-simplices $C_m=C_m(X^i)$are explicitly defined 
by one of the vertices $X^i=(X,Y,Z)$ and four of the vertices 
(\ref{threetwo}) as follows:
\begin{mathletters}
\label{threefour}
\begin{eqnarray}
C_1&=&\{j-1,p+j-1,(k+1)p+j-1,(k+2)p+j-1,X^i\} \label{threefoura}\ , \\ 
C_2&=&\{j-1,p+j,(k+1)(p+1)+j-1,(k+2)(p+1)+j-1,X^i\}\label{threefourb}\ , 
\\ C_3 &=&\{j-1,j,k+j,j+k+1,X^i\}\ (mod\ n)\ , \qquad j=1,2,...,n\ , \ 
k=1,2,...,p-1.
\label{threefourc}
\end{eqnarray}
\end{mathletters}
For each $p$, the triangulation $CP^2_{n+3}$ is given by the union of the 
three balls
$B_i=C_j(X^i)\cup C_k(X^i), \quad i=1,2,3,\quad j,k\neq i$. 

The number of 4-simplices in the triangulation $CP^2_{n+3}$ is 
$n_4=6(p^3-1)$.
One may employ the necessary relation for a combinatorial 4-manifold 
\begin{equation}
5n_4=2n_3
\label{threesix}
\end{equation}
together with the Euler number
\begin{equation}
n_0-n_1+n_2-n_3+n_4=3
\label{threeseven}
\end{equation}
and the Dehn-Sommerville relation
\begin{equation}
2n_1-3n_2+4n_3-5n_4=0
\label{threeeight}
\end{equation}
to compute the number $n_k$ of $k$-simplices. The numbers are $n_0=n+3,\ 
n_1=3\,p\,n,\ n_2=2\, (6p-5) \, n,\ n_3=15\, (p-1)\, n,\ n_4=6 \, 
(p^3-1)\ .$
\vskip .13 in
\centerline{\bf Table I: The Orbits of Triangles for $p=2$} $$
\vbox{\halign{#\hfill&\quad#\hfill&\quad#\hfill&\quad 
#\hfill&\quad#\hfill&\quad#\hfill \cr
\noalign{\vskip 1pt\hrule\vskip 1pt}
\noalign{\vskip 1pt\hrule\vskip 1pt}
\cr
\hfill Triangle\hfill &\hfill Are All\hfill & Number of & & Number of \cr

\hfill Orbit\hfill &\hfill Edges on\hfill & Vertices on &\hfill 
Length\hfill
& \hfill Incident\hfill
& Representative\cr
& \hfill$T^2$?\hfill &\hfill$T^2$\hfill & & 4-Simplices & \cr 
\noalign{\vskip 1pt\hrule\vskip 1pt}
\cr
\hfill$\alpha$\hfill & \hfill yes\hfill &\hfill 3\hfill &\hfill 14\hfill 
&\hfill 6\hfill &~~~~~(0,1,3)\cr
\hfill$\beta$\hfill &\hfill no\hfill &\hfill 3\hfill &\hfill 21\hfill 
&\hfill 4\hfill &~~~~~(0,1,4)\hfill \cr
\hfill$\gamma$\hfill &\hfill no \hfill &\hfill 2\hfill &\hfill 21\hfill 
&\hfill 5\hfill &~~~~~(0,1,7)\hfill \cr
\hfill$\delta$\hfill &\hfill no\hfill &\hfill 2\hfill &\hfill 21\hfill 
&\hfill 4\hfill &~~~~~(0,1,8)\hfill \cr
\hfill$\epsilon$\hfill &\hfill no\hfill &\hfill 2\hfill &\hfill 21\hfill 
&\hfill 3\hfill &~~~~~(0,1,9)\hfill \cr
\cr
\noalign{\vskip 1pt\hrule\vskip 1pt}
\noalign{\vskip 1pt\hrule\vskip 1pt}
}}
$$
The symmetry group of $CP^2_{n+3}$ is generated by the cyclic 
permutations
\begin{eqnarray}
T&=&(k\rightarrow k+1\ (mod\, n)),\nonumber\\ S&=&(k\rightarrow pk \ 
(mod\, n))\ (XYZ)\ ,\nonumber \\ R&=&(k\rightarrow -k\ (mod\, n))\ .
\label{threefive}
\end{eqnarray}
We denote the group by $G_p$.

When the symmetry group of the triangulation is an isometry of the 
simplicial geometry we say that we have a {\it maximally symmetric} 
geometry. We shall be interested only in maximally symmetric solutions of 
the Regge equations (1.1).
As a consequence of this isometry, the edges of a maximally symmetric 
geometry lying on $T^2$ are all equal,
and the length will be taken to be $a$. The cone edges from the vertices 
in $T^2$ to the ending vertices $X,Y$ or $Z$ are also all equal. The 
length of these will be taken to be $b$. There exist $p-2$ trajectories 
of internal edges among the vertices of $T^2$ for $p>2$. These
will have the respective lengths $c_k, \quad k=1,2,...,p-2$. 

The $2n(6p-5)$ triangles are of three distinct types. Those with the 
edges lying on the boundary torus form a single orbit of length $2n$ 
under the isometries. Two orbits, from representative triangles 
$(0,1,p+1)$ and $(0,p,p+1)$, are generated by the translations $T$ and 
interchanged by the reflections $R$. For $p\ge3$, there exist $p-2$ 
orbits of interior triangles, each of length $3n$. The interior triangles 
have three vertices and two edges on the boundary torus and one edge 
inside a flat solid torus. The third type of triangle lies on a side of a 
cone, with one of the vertices being either $X$, $Y$ or $Z$. The cone 
triangles form a single orbit of length $3n$ under the symmetries. There 
are $2p-2$ orbits of the four-simplices defined by (\ref{threefour}).

\section{Solutions and Non-Solutions}
\label{sec:IV}

In this Section we consider solutions to the Regge equations 
(\ref{onetwo}) on the triangulations $CP^2_{n+3}$. These equations define 
an extremum of the Regge action $I(s_i, M)$ that is the simplicial analog 
of the Euclidean continuum action \begin{equation}
\ell^2 I[g,M] = - \int_M d^4 x \sqrt{g}\ (R-2\Lambda) \label{fourone}
\end{equation}
\vskip .13 in
\centerline{\epsfxsize=4.75truein \epsfbox{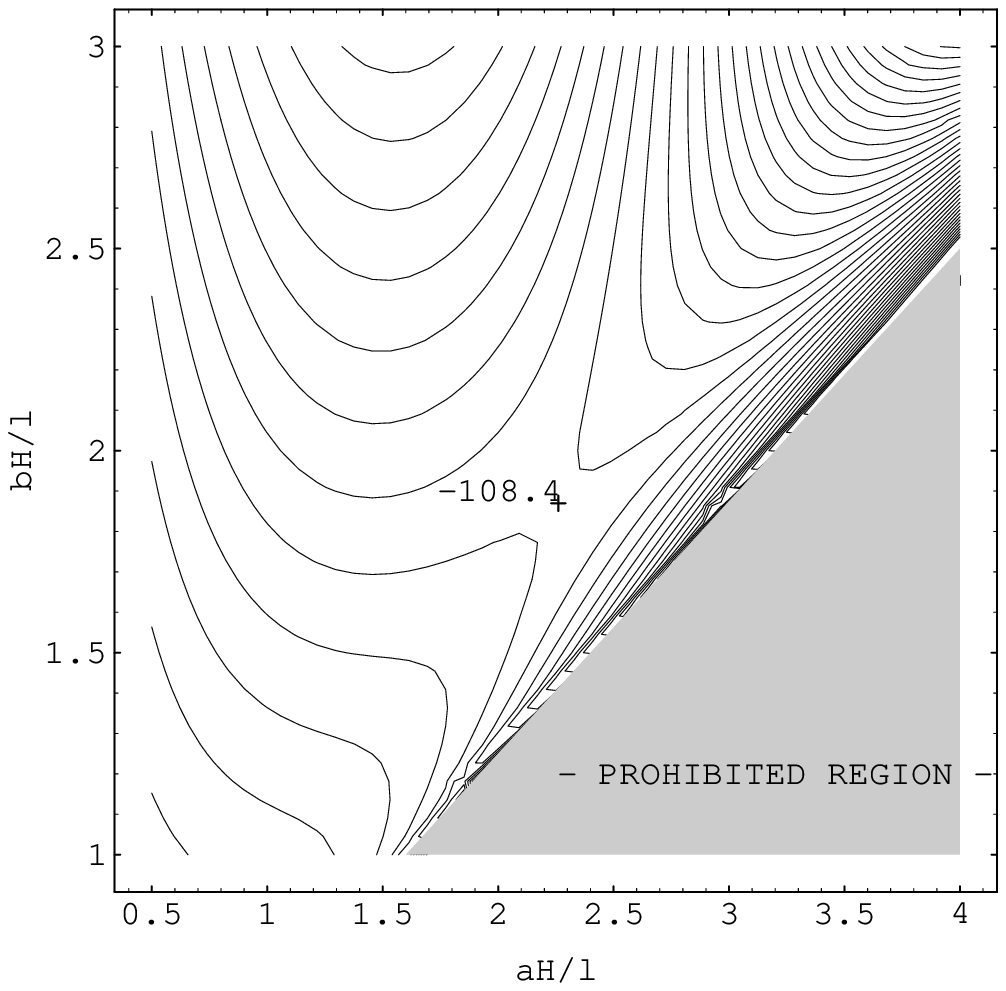}} \vskip .13 in
\begin{quote}
{\protect\small { \sl Figure 1. A contour map of the action for the the 
$p=2$ triangulation $CP^2_{10}$. Contour lines of constant $H^2I$, where 
$I$ is the action (4.2), are shown as a function of the two independent 
edge lengths $a$ and $b$ in a maximally symmetric assignment of edge 
lengths. The contour lines of $H^2 I$ are spaced by $2.35$ dimensionless 
units.
One or more of the triangle inequalities or their higher dimensional 
analogs are violated in the ``prohibited region'' at the lower right of 
the diagram. There is a saddle point and a solution of the Regge 
equations at $a=2.26 l/H$ and $b=1.87 l/H$.}} \end{quote}
Here, $\ell$ is $\sqrt{16\pi}$ times the Planck length in units where 
$\hbar=c=1$ and $\Lambda$ is the cosmological constant. Specifically, the 
Regge action is, for a closed manifold,
\begin{equation}
\ell^2 I (s_i,M) = -2 \sum\limits_{\sigma\in \Sigma_2} V_2 (\sigma) 
\theta (\sigma) + \left(6H^2/\ell^2\right) \sum\limits_{\tau\in\Sigma_4} 
V_4 (\tau) \ . \label{fourtwo}
\end{equation}
The first sum is over all triangles and defines the curvature action 
$\ell^2 I_{\rm curv}$. The area of triangle $\sigma$ is $V_2(\sigma)$ and 
$\theta(\sigma)$ is its deficit angle which is $2\pi$ minus the sum of 
the dihedral angles between the tetrahedral faces of the four-simplices 
that meet $\sigma$. The second sum is over the four-simplices $\tau$. 
$V_4(\tau)$ is the volume of the four-simplex $\tau$, and we have 
introduced the dimensionless measure of the cosmological constant $H^2$ 
by writing $\Lambda = 3H^2/\ell^2$. For more details on the meaning of 
these quantities as well as an explanation of how to express them as 
functions of the squared edge-lengths $s_i$ see, {\it e.g.} Paper I. 
\eject

\centerline{\bf Table II: Areas and Deficit Angles of the Triangles} 
$$\vbox{\halign{\hfill#\hfill&\quad\hfill#\hfill&\quad\hfill#\hfill\cr 
\noalign{\vskip 1pt\hrule\vskip 1pt}
\noalign{\vskip 1pt\hrule\vskip 1pt}
\cr
Triangle& Area& Deficit\cr
Orbit &	& Angle \cr
\noalign{\vskip 1pt\hrule\vskip 1pt}
\cr
$\alpha$ & $A_1$ & $2\pi - 6\theta_1$ \cr $\beta$ & $A_1$ & $2\pi - 
4\theta_1$\cr
$\gamma$ & $A_2$ & $2\pi - 5\theta_2$\cr $\delta$ & $A_2$ & $2\pi - 
4\theta_2$\cr $\epsilon$ & $A_2$ & $2\pi - 3\theta_2$\cr \cr
\noalign{\vskip 1pt\hrule\vskip 1pt}
\cr
&$A_1 = \left(\sqrt{3}/4\right) a^2 \qquad A_2 = \frac{1}{4} a \left(4b^2 
-
a^2\right)^{\frac{1}{2}}$\cr
&$\theta_1 = \arccos \left\{\sqrt{2a}/\left[4(3b^2 - 
a^2)^{\frac{1}{2}}\right]\right\}$\cr
&$\theta_2 = \arccos \left[(a^2 - 2b^2)/(2a^2 - 6b^2)\right]$\cr \cr
\noalign{\vskip 1pt\hrule\vskip 1pt}
\noalign{\vskip 1pt\hrule\vskip 1pt}
}}
$$
We seek only maximally symmetric solutions of the Regge equations 
(\ref{oneone}) as described in Section III. The simplest case is $p=2$. 
The symmetry group $G_2$ is of order 42. Maximally symmetric geometries 
are characterized by the two independent edge lengths $a$ and $b$. 
Thereare five orbits for the triangles under $G_2$ whose properties are 
spelled
out in Table I. Their
areas and deficit angles are given in
Table II in terms of $a$ and $b$.
There are 2 orbits of length 21 for
the four-simplices. In a maximally symmetric geometry all volumes have 
the same value:
\begin{equation}
V_4={1\over 4}a^3(8b^2-3a^2)^{1/2} \ . \label{fourthreea} \end{equation}
We developed a Mathematica program\footnote{This program is available 
from Z.~Perj\'es} to find various properties including the analytic form 
of the action (\ref{fourtwo})
for an arbitrary triangulated spacetime in Regge calculus. The analytic 
expression for the action when $p=2$ is \begin{eqnarray}
\ell^2 I=&
{21\,a^3\,\sqrt{ -3\, a^2 +
8\, b^2 }\, H^2\over 8\, l^2} -
35\,\sqrt{3}\, a^2\,\pi -
63\,\sqrt{- a^4 + 4\, a^2\, b^2}\,\pi\nonumber\\ & +
84\,\sqrt{3}\, a^2\,
\arccos ({ a\over
2\,\sqrt{2}\,\sqrt{\,
- a^2 + 3\, b^2 }})\nonumber\\
& +
126\,\sqrt{- a^4 + 4\, a^2\, b^2}\,
\arccos ({\, - a^2 + 2\, b^2
\over
2\,\left( a^2 -
3\, b^2 \right)}) \ .
\label{fourthree}
\end{eqnarray}

For $p=3$ the maximally symmetric geometries are specified by three 
independent edge-lengths $a, b$ and $c$. The analytic expression for the 
Regge action is
\begin{eqnarray}
\ell^2 I&={{{39\over8}{H^2\over
l^2}\Bigl(\sqrt{c^4(-2a^2+c^2)(2a^2-8b^2+c^2)}+ 
\sqrt{a^2c^2(-4a^4+12a^2b^2+a^2c^2-4b^2c^2)}\Bigr)}}\nonumber\\ 
&-13\Biggl(2\sqrt{3}a^2\pi+9\sqrt{-a^4+4a^2b^2}\pi-6\sqrt{3}a^2\arccos 
\biggl({{ac}\over{2\sqrt{(a^2-3b^2)(-3a^2+c^2)}}}\biggr)\nonumber\\ 
&-3\sqrt{-a^4+4a^2b^2}\arccos\biggl({{-2a^4+6a^2b^2+a^2c^2-4b^2c^2}\over 
{2(a^4-3a^2b^2)}}\biggr)\nonumber\\
&-12\sqrt{-a^4+4a^2b^2}\arccos\biggl({{(-a^2+2b^2)c}\over 
{2\sqrt{(a^2-3b^2)A}}}\biggr)\nonumber\\
&-12\sqrt{-a^4+4a^2b^2}\arccos\biggl({{(-a^2+2b^2)c^2}\over 
{2A}}\biggr)+9\pi\sqrt{4a^2c^2-c^4}\nonumber\\ 
&-12\arccos\biggl({{c(2a^2-c^2)}\over
{2\sqrt{2}\sqrt{(-2a^2+c^2)A}}}\biggr)\sqrt{4a^2c^2-c^4}\nonumber\\ 
&-6\arccos\biggl({{a^4-a^2c^2/2}\over
{\sqrt{a^2(-3a^2+c^2)A}}}\biggr)
\sqrt{4a^2c^2-c^4}+6\pi C\nonumber\\
&-3\arccos\biggl({{(a^2-2b^2)(-2a^2+c^2)}\over{2A}}\biggr) 
C\nonumber\\
&-6\arccos\biggl({{-2a^4+8a^2b^2-6b^2c^2+c^4}\over 
{2A}}\biggr)C\Biggr)\ ,
\label{fourfour}
\end{eqnarray}
where
\begin{equation}
A=\sqrt{a^4-4a^2b^2+b^2c^2},\qquad C=\sqrt{4b^2c^2-c^4} \ . 
\label{fourfive}
\end{equation}
And so on.

We now consider the solutions of the Regge equations (\ref{onetwo}) for 
$p=2$ and $p=3$. For maximally symmetric solutions it is enough to 
extremize the actions (\ref{fourthree}) and (\ref{fourfour}) as a 
function of the independent squared edge lengths. We searched for 
solutions by a
combination of graphical and numerical techniques --- using contour maps 
of the action to understand its behavior and then numerical evaluations 
of the Regge equations to find solutions. We studied the norm $N(s_i) 
\equiv \Sigma_i (\partial I/\partial s_i)^2$ to locate solutions 
accurately at its zeros. We compared results following from the analytic 
forms (\ref{fourthree}) and (\ref{fourfour}) with a general code which 
computes each of the Regge equations $\partial I/\partial s_i\ , i=1, 
\cdots, n_1$ without assumption of symmetry. There
was agreement.

A contour map of the action of
the $p=2$ triangulation $CP^2_{10}$ is shown in {\it Figure 1}. Using the 
methods described above we found we found a maximally symmetric
solution at
\begin{equation}
a= 2.26(\ell/H)\quad , \quad b=1.87(\ell/H) \ . \label{foursix}
\end{equation}
The properties of this solution can be compared with the maximally 
symmetric continuum metric --- the Fubini-Study metric described in 
Section II --- and with the solution on the K\"uhnel-Lassmann 
triangulation ($CP^2_9$) of $CP^2$ discussed in Paper II. To do that we 
compare
the
values of two invariants among the different solutions. One is the total 
volume $V_{\rm tot}$ and the other is the curvature action defined by 
\begin{equation}
V_{\rm tot} = \sum\limits_{\tau\in\Sigma_4} V_4(\tau) \equiv {\cal 
B}\left(\ell/H\right)^4 \ ,
\label{fourseven}
\end{equation}
\begin{equation}
I_{\rm curv} = -2\sum_{\sigma\in \Sigma_2} V_2 (\sigma) \theta (\sigma) 
\equiv - {\cal A}\left(V_{\rm tot}/\ell^4\right)^{\frac{1}{2}} \ .
\label{foureight}
\end{equation}
\centerline{\bf Table III: Comparison of Solutions} $$
\vbox{\halign{\hfill#\hfill&\quad\hfill#\hfill&\quad\hfill#\hfill\cr 
\noalign{\vskip 1pt\hrule\vskip 1pt}
\noalign{\vskip 1pt\hrule\vskip 1pt}
\cr
& Curvature Action & Total Volume\cr
&$\cal A$& $\cal B $\cr
\cr
\noalign{\vskip 1pt\hrule\vskip 1pt}
\cr
$CP^2_9$ & 50. & 18. \cr
$CP^2_{10}$ & 51. & 18. \cr
Fubini-Study & 53.29. & 19.74. \cr
\cr
\noalign{\vskip 1pt\hrule\vskip 1pt}
\noalign{\vskip 1pt\hrule\vskip 1pt}
}}
$$
Table III shows the dimensionless constants $\cal A$ and $\cal B$ for the 
various
solutions.\footnote{In Paper II, the constant $\cal A$ was denoted by $a$ 
in eq.(3.4) and Table II. We use upper case here to avoid confusion with 
the edge-length $a$. The value of $\cal A$ for the 
Fubini-Study metric was quoted incorrectly in Table II of Paper II. Its 
correct
value is in eq.(4.10) of this paper.} For the continuum Fubini-Study 
metric we have in particular
\begin{equation}
{\cal A}= 12\sqrt{2}\ \pi \quad , \quad {\cal B}= 2\pi^2 \ . 
\label{fournine}
\end{equation}
By these measures the progression from $CP^2_9$ to $CP^2_{10}$ is only 
slight and both give tolerable approximations to the continuum results. 

For $p=3$ we found no solution. A careful search of the norm $N(s_i) = 
\Sigma_i (\partial I/\partial s_i)^2$ showed it approaching a minimum 
value but that value was not zero. We do not find a saddle point for the 
triangulation $CP^2_{16}$. Moving from $p=2$ to $p=3$ therefore results 
in a worse approximation to the continuum Fubini-Study metric. 

\section{Eigenvalues of the Second Variation} \label{sec:V}

In addition to the extrema of the action, its second variation 
\begin{equation}
I^{(2)}_{ij} = \frac{\partial^2 I}{\partial s_i \partial s_j} 
\label{fiveone}
\end{equation}
is important for the evaluation of the semiclassical approximation to 
sums over simplicial geometries like (\ref{oneone}). The prefactor of the 
semiclassical approximation is the inverse square root of the determinant 
of $I_{ij}^{(2)}$ evaluated at the contributing extrema. This determinant 
is
the product of the eigenvalues of $I^{(2)}_{ij}$. The signs of these 
eigenvalues determine the steepest 
descents directions at
the extrema. Small eigenvalues signal the approximate diffeomorphism 
invariance that occurs in Regge calculus \cite{Har85a}. For these and 
other reasons the eigenvalues of (\ref{fiveone}) at the extrema of the 
action are of interest.
The eigenvalues of $I^{(2)}_{ij}$ for the $p=2$ symmetric solution are 
shown in Table IV. The table shows the eigenvalues $\lambda$ and their 
multiplicities $\rho_\lambda$. The eigenvalues of a maximally symmetric 
solution may be classified by the
irreducible representations of that group and the multiplicities of these 
eigenvalues are given by the dimensions of these irreducible 
representations.
\vskip .13 in
\centerline{\bf Table IV: Eigenvalues and Multiplicities of 
$\partial^2I/\partial s_i
\partial s_j$}
\centerline{\bf Evaluated at the $p=2$ Stationary Point} $$
\vbox{\halign{\hfill#\hfill&\qquad\qquad\hfill#\hfill&\qquad\qquad 
\hfill#\hfill&\qquad\qquad
\hfill#\hfill\cr
\noalign{\vskip 1pt\hrule\vskip 1pt}
\noalign{\vskip 1pt\hrule\vskip 1pt}
\cr
$\ell^4 H^{2} \lambda$ & $\rho_\lambda$ & $\ell^4H^{2}\lambda$ & 
$\rho_\lambda$\cr
\cr
\noalign{\vskip 1pt\hrule\vskip 1pt}
\cr
-1.74 & 1 & -.12 & 6 \cr
-1.24 & 2 & -.05 & 6 \cr
-~.63 & 6 & +.07 & 6 \cr
-~.39 & 6 & +.28 & 1 \cr
-~.18 & 8 &	& \cr
\cr
\noalign{\vskip 1pt\hrule\vskip 1pt}
\noalign{\vskip 1pt\hrule\vskip 1pt}
}}
$$
There are seven classes of the group $G_2$ whose orders and characters 
are given in Table V. This shows the classes arranged horizontally with 
their orders on the top line and the value of the characters on those 
classes arranged vertically for the seven possible irreducible 
representations.
An element of the symmetry group sends some edges into others thereby 
producing a 42 dimensional reducible representation of $G_{42}$. The 
irreducible representations it contains label the eigenvalues of 
$I_{ij}$. The number of times a given irreducible representation occurs 
in this reducible representation may be found by analyzing its characters 
into the characters of the irreducible representations. (For more 
explicit details see {\it e.g.}~Paper II, Section IV.) The result is that 
the reducible
representation decomposes as $2\cdot 1_1+0\cdot 1_2+0\cdot1_3+0\cdot 1_4 
+ 2\cdot 1_5 +2\cdot 1_6 +6\cdot 6_1$. Here, the individual terms in this 
sum correspond to the
the irreducible representations in Table II and occur in the order listed 
there.
The second factor in each summand is the dimension of the irreducible 
representation and multiplication indicates the number of times it occurs 
in the 42-dimensional reducible representation. The
multiplicities of the eigenvalues calculated numerically and shown in 
Table II are consistent with this analysis although there is an 
unexplained degeneracy between a 6 and two 1's resulting in the 
multiplicity of 8 and between two 1's resulting in a multiplicity of 2. 
This kind of degeneracy also occurs with the eigenvalues of the symmetric 
solution of $CP^2_9$ (Paper II) in the flat, symmetric assignment of edge 
lengths to certain triangulations of $T^3$ \cite{HMWup}. It thus seems to 
be a feature of a wide range of triangulations. 

Other features of the eigenvalues are immediately apparent from Table IV. 
Two of the eigenvalues are positive and the rest are negative. The 
largest positive eigenvalue corresponds to a uniform change in all edges. 
The number of positive eigenvalues available to represent
true physical degrees of freedom is thus slightly larger than in the 
slightly smaller triangulation, $CP^2_9$, discussed in Paper II. 

Some eigenvalues are small but none are exactly zero. There are thus 
directions in the space of edge lengths in which the action varies slowly 
as it would for an approximate diffeomorphism but none in which it is 
exactly constant.
\vfill\eject
\centerline{\bf Table V: Character Table for the Symmetry Group of the 
$p=2$
Triangulation}
$$
\vbox{\halign{\hfill#\hfill&\qquad\qquad\hfill#\hfill&\qquad 
\hfill#\hfill&\qquad\qquad\hfill#\hfill&\qquad\qquad 
\hfill#\hfill&\qquad\qquad\hfill#\hfill&\qquad\qquad\hfill 
#\hfill&\qquad\qquad\hfill#\hfill\cr
\noalign{\vskip 1pt\hrule\vskip 1pt}
\noalign{\vskip 1pt\hrule\vskip 1pt}
\cr
1 & 6 & 7 & 7 & 7 & 7 & 7 \cr
\cr
\noalign{\vskip 1pt\hrule\vskip 1pt}
\cr
1 & 1 & 1 & 1 & 1 & 1 & 1 \cr
1 & 1 & -1 & 1 & 1 & -1 & -1 \cr
1 & 1 & -1 & $\epsilon$ & $\epsilon^2$ & $-\epsilon$ & $-\epsilon^2$ \cr 
1 & 1 & -1 & $\epsilon^2$ & $\epsilon$ & $-\epsilon^2$ & $-\epsilon$ \cr 
1 & 1 & 1 & $\epsilon$ & $\epsilon^2$ & $\epsilon$ & $\epsilon^2$ \cr 1 & 
1 & 1 & $\epsilon^2$ & $\epsilon$ & $\epsilon^2$ & $\epsilon$ \cr 6 & -1 
& 0 & 0 & 0 & 0& 0 \cr
\cr
\noalign{\vskip 1pt\hrule\vskip 1pt}
\cr
& $\epsilon = \exp(2\pi i/3)$ \cr
\cr
\noalign{\vskip 1pt\hrule\vskip 1pt}
\noalign{\vskip 1pt\hrule\vskip 1pt}
}}
$$

\section{Discussion}
\label{sec:VI}

One might have naively expected that the more refined the triangulation 
of a manifold $M$ is, the better the approximation the solution to it's
Regge equations will be to the solution in the continuum. The 
Banchoff-K\"uhnel sequence $CP^2_{n+3}$ shows this is not generally the 
case. In the first two steps one moves from a solution which is a 
reasonable approximation by some measures to no solution at all! The 
reason for this is to be found in the nature of the refinement. As we 
take triangulations of higher $p$ it is the triangulation of torus $T^2$ 
which is refined, while the number of ``exterior'' vertices $(X, Y, Z)$ 
remain fixed. If the vertices are embedded in the Fubini-Study geometry 
as described in \cite{BK92}, the edges in $T^2$ better approximate the 
Fubini-Study distances between vertices with increasing $p$. However, the 
cone edges $b$ is constant.

This example shows that sequences of triangulations of a given manifold 
will have to be chosen carefully if the solutions of the Regge equations 
are to converge to the continuum solution of the Einstein equations. 

\acknowledgements

This work was supported in part by NSF grants PHY 95-07065, PHY94-07194, 
and OTKA grant T 017176. Z.P.~acknowledges
the support of a U.S.~Government Fulbright Research Fellowship.

\end{document}